\documentclass[12pt,3p,dvipdfmx]{elsarticle}
\usepackage{amsmath,amssymb,amsthm,graphicx,psfrag,bm}
\usepackage{layout}
\usepackage{color}
\usepackage{makeidx}
\usepackage{multicol,multienum}
\usepackage{enumerate}
\usepackage{tabularx}
\biboptions{sort&compress}
\def\Fdagger#1{\setbox0=\hbox{$#1$}             
   \dimen0=\wd0                                 
   \setbox1=\hbox{/} \dimen1=\wd1               
   \ifdim\dimen0>\dimen1                        
      \rlap{\hbox to \dimen0{\hfil/\hfil}}      
      #1                                        
   \else                                        
      \rlap{\hbox to \dimen1{\hfil$#1$\hfil}}   
      /                                         
   \fi}

\title{
Radiative neutrino mass, dark matter and electroweak baryogenesis 
from the supersymmetric gauge theory with confinement
}
\author[toyama]{Shinya Kanemura}
\author[toyama]{Naoki Machida}
\author[kogakuin]{Tetsuo Shindou}
\address[toyama]{Department of Physics, University of Toyama, \\
3190 Gofuku, Toyama, 930-8555, Japan}
\address[kogakuin]{Division of Liberal-Arts, Kogakuin University,\\
1-24-2 Nishi-Shinjuku, Tokyo, 163-8677, Japan}

\begin{document}
\begin{abstract}
We propose a simple model to explain neutrino mass, dark matter and
baryogenesis based on the extended Higgs sector which appears 
in the low-energy effective theory of a supersymmetric gauge theory 
with confinement.
We here consider the SU(2)$_H$ gauge symmetry with three flavours
of fundamental representations which are charged under the standard 
SU(3)$_C\times$ SU(2)$_L\times$U(1)$_Y$ symmetry
and a new discrete $Z_2$ symmetry.
We also introduce $Z_2$-odd right-handed neutrino superfields
in addition to the standard model matter superfields. 
The low-energy effective theory below the confinement scale
contains the Higgs sector with fifteen composite superfields,
some of which are $Z_2$-odd.
When the confinement scale is of the order of ten TeV,
electroweak phase transition can be sufficiently of first order,
which is required for successful electroweak baryogenesis.
The lightest $Z_2$-odd particle can be a new candidate for dark matter, 
in addition to the lightest $R$-parity odd particle.
Neutrino masses and mixings can be explained by the quantum effects of 
$Z_2$-odd fields via the one-loop and three-loop diagrams.
We find a benchmark scenario of the model, where all the constraints from 
the current neutrino, dark matter, lepton flavour violation and LHC data
are satisfied.
Predictions of the model are shortly discussed.
\end{abstract}
\vbox{
    \halign{#\hfil        \cr
           UT-HET-092    \cr
		   KU-PH-014
           \cr}}
\vspace*{10mm}
\maketitle

\section{Introduction}
The Higgs boson has been discovered at the LHC, and 
its measured properties are currently consistent 
with the standard model (SM)\cite{HiggsLHC}.
However, the minimal Higgs sector in the SM is just an assumption.
We still do not know the essence of the Higgs boson and 
the structure of the Higgs sector.
Is the Higgs boson really a scalar particle or otherwise a composite
state?
What is the fundamental physics behind the Higgs dynamics?
What is the origin of vacuum condensation?
How many Higgs fields are there?
Answers for these questions directly correspond to the 
paradigm of the fundamental theory beyond the SM.
At the same time, the possibility of various extended Higgs sectors
provides us an idea that 
the Higgs sector would be strongly related to the phenomena such as 
tiny neutrino masses and mixing\cite{nuexp}, 
the existence of dark matter (DM) \cite{WMAP} and the 
baryon asymmetry of the Universe (BAU)\cite{WMAP}, none of which can be 
explained in the SM.

Among several possibilities for baryogenesis\cite{HighTBG,EWBG}, 
there is a scenario 
so-called electroweak baryogenesis\cite{EWBG},
where the BAU could be explained by the 
dynamics of the Higgs potential
when the electroweak phase transition is of strongly first order. 
It is well-known that the electroweak baryogenesis cannot be 
realized within the SM.
Hence, a non-minimal Higgs sector has to be introduced for the successful 
scenario of electroweak baryogenesis\cite{extendedHiggsforEWBG,SUSYEWBG,KanemuraSenahaShindou}. 
With the discovered Higgs boson mass to be 126 GeV, 
the condition of the strong first-order phase transition (1stOPT) requires 
at least one of the self-coupling constants in the Higgs potential
to be relatively large.
A phenomenological consequence of the theory with the strong 
1stOPT is a significantly larger
triple Higgs boson coupling than the SM prediction,
by which the scenario of the electroweak baryogenesis can be 
tested at future collider experiments\cite{KanemuraOkadaSenaha}.
At the same time, such a large self-coupling constant in the Higgs potential 
tends to cause early brow-up of the running coupling constant, and 
the Landau pole\cite{Landaupole} can appear at the scale much below the Planck 
scale\cite{KanemuraSenahaShindouYamada}.
In this case, the ultraviolet picture above the Landau pole should
also be considered\cite{KanemuraShindouYamada}.

One possible explanation for tiny neutrino masses is 
based on the seesaw mechanism, where neutrino masses are 
explained at the tree level with introducing very heavy 
right-handed (RH) neutrinos\cite{seesaw1}, 
Higgs triplet fields\cite{seesaw2} or
fermion triplet fields\cite{seesaw3}.
An alternative idea is to generate tiny neutrino masses 
radiatively by introducing extended Higgs sectors at the 
TeV scale.
Since the original model was proposed by A.~Zee\cite{Zee},
many models\cite{zeebabu,SUSYzeebabu,knt,Ma,AKS,AKS2} have been proposed along this line.
In a class of models where neutrino masses are generated 
radiatively at loop levels, an unbroken discrete
$Z_2$ symmetry and RH neutrinos are introduced
such that the RH neutrinos have the 
odd quantum number to make neutrino Yukawa coupling constants 
absent at the tree level\cite{knt,Ma,AKS,AKS2}.
The same symmetry also guarantees the stability of the lightest 
$Z_2$-odd particle, so that it can be a DM candidate. 
The model proposed by E.~Ma (the Ma model) is the simplest model of 
this category\cite{Ma} where the neutrino masses are generated at the 
one-loop level, and in the model proposed in Ref.~\cite{AKS,AKS2} 
(the AKS model) they are generated at the three-loop level.
Both models have DM candidates. Furthermore, in the AKS model, 
the strong 1stOPT is also realized.
Although these models are phenomenologically acceptable, additional 
scalar particles are introduced in an {\it ad-hoc} way which seems
rather artificial.
Fundamental theories are desirable in which these phenomenological 
models are deduced in the low-energy effective theory.

In this Letter, 
we propose a simple model 
whose low-energy effective theory 
can explain neutrino mass, DM and baryogenesis.
In this model, the supersymmetric (SUSY) extended Higgs sector appears in 
the low-energy effective theory of a SUSY gauge theory 
with confinement\cite{Intriligator:1995au}.
With an additional $Z_2$ symmetry,
all the scalar fields introduced in the Ma model and 
the AKS model automatically appear, so that 
introducing a RH neutrino superfield with the odd quantum number,
neutrino mass, DM and baryogenesis can be explained simultaneously 
by a hybrid mechanism of the Ma model and the AKS model in the framework of SUSY.
Consequently there are two kinds of the DM candidates: {\it i.e.}, one comes from the lightest 
$R$-parity odd SUSY particle, and the other is the lightest $Z_2$-odd particle,
so that the DM scenario of our model is multi-component DM scenario.

We introduce the SU(2)$_H$ gauge symmetry with three flavours
of fundamental 
representations \cite{fatHiggs,KanemuraShindouYamada}, 
which are charged under the standard 
SU(3)$_C\times$ SU(2)$_L\times$U(1)$_Y$ symmetry
and a new discrete $Z_2$ symmetry.
In addition to the SM matter superfields,
we also introduce $Z_2$-odd RH neutrino superfields.
Then the low-energy effective theory below the confinement scale
contains the Higgs sector with fifteen composite superfields,
some of which are $Z_2$-odd.
Electroweak phase transition can be of sufficiently strong first-order,
when the confinement scale is of the order of ten TeV 
\cite{KanemuraSenahaShindouYamada,KanemuraMachidaShindouYamada}.
In addition to the lightest $R$-parity odd particle,
the lightest $Z_2$-odd particle can be a new candidate for DM.
We can explain 
neutrino masses and mixings 
by the quantum effects of $Z_2$-odd fields via the one-loop and 
three-loop diagrams.

We find a benchmark scenario of the model, where all the constraints from 
the current neutrino\cite{nuexp}, DM\cite{WMAP,LUX,XENON100}, 
lepton flavour violation (LFV) \cite{MEG} and LHC data
are satisfied.
We also comment on predictions of the model.

\section{The SUSY gauge theory with confinement and its low-energy effective theory}
Our model is based on a SUSY model with the $\text{SU}(2)_H\times Z_2$ symmetry. 
We introduce six chiral superfields, $T_i$ $(i=1,\cdots,6)$, which are doublet under the $\text{SU}(2)_H$ 
gauge symmetry.
The chiral superfields $T_i$'s also have gauge quantum number under the SM 
gauge symmetry, $\text{SU}(3)_C\times \text{SU}(2)_L\times \text{U}(1)_Y$, and 
moreover 
quantum numbers of the $Z_2$ parity are assigned.
In addition, a RH neutrino superfield $N_R^c$ is also introduced. 
As similar to the setup proposed in Ref.~\cite{KanemuraMachidaShindouYamada},
this is a singlet chiral superfield for both the SU(2)$_H$ and the SM gauge symmetry 
but it has an odd parity under the $Z_2$ symmetry.
The SM charges and the $Z_2$ parity assignments on $T_i$'s and $N_R^c$ are shown in Table~\ref{FieldContent}.

\begin{table}[t]
	\caption{The SM charges and the $Z_2$ parity assignment on the 
		SU(2)$_H$ doublets $T_i$ and the SU(2)$_H$ singlet 
	RH neutrino $N_R^c$.\label{FieldContent}}
		\begin{center}
			\begin{tabular}{|c|c|c|c|c|c|}\hline
				Superfield&SU(2)$_H$&SU(3)$_C$&SU(2)$_L$&U(1)$_Y$&$Z_2$\\ \hline
				$\displaystyle
				\begin{pmatrix}
					T_1\\
					T_2
				\end{pmatrix}$& 2&1&2&0&$+1$\\ \hline
					$T_3$&2&1&1&$+1/2$&$+1$\\ \hline
					$T_4$&2&1&1&$-1/2$&$+1$\\ \hline
					$T_5$&2&1&1&$+1/2$&$-1$\\ \hline
					$T_6$&2&1&1&$-1/2$&$-1$\\ \hline \hline
					$N_R^c$&1&1&1&0&$-1$\\ \hline
			\end{tabular}
		\end{center}
\end{table}

\begin{table}[t]
	\caption{The field contents of the Higgs sector below the confinement scale $\Lambda_H$.\label{FieldContentHiggs}}
	\begin{center}
		\renewcommand{\arraystretch}{1.3}
		\begin{tabular}{|c|c|c|c|c|c|}\hline
		Superfield&SU(3)$_C$&SU(2)$_L$&U(1)$_Y$&$Z_2$\\ \hline
			$H_d\equiv\begin{pmatrix}
				H_{14}\\
				H_{24}\\
			\end{pmatrix}$
			&1&2&$-1/2$&$+1$\\ \hline
			$H_u\equiv\begin{pmatrix}
				H_{13}\\
				H_{23}\\
			\end{pmatrix}$
			&1&2&$+1/2$&$+1$\\ \hline
			$\Phi_d\equiv\begin{pmatrix}
				H_{15}\\
				H_{25}\\
			\end{pmatrix}$
			&1&2&$-1/2$&$-1$\\ \hline
			$\Phi_u\equiv\begin{pmatrix}
				H_{16}\\
				H_{26}\\
			\end{pmatrix}$
			&1&2&$+1/2$&$-1$\\ \hline
			$\Omega_-\equiv H_{46}$&1&1&$-1$&$-1$\\ \hline
			$\Omega_+\equiv H_{35}$&1&1&$+1$&$-1$\\ \hline
			$N\equiv H_{56},N_{\Phi}\equiv H_{34},N_{\Omega}=H_{12}$&1&1&$0$&$+1$\\ \hline
			$\zeta \equiv H_{36},\eta\equiv H_{45}$&1&1&$0$&$-1$\\ \hline
		\end{tabular}
	\end{center}
\end{table}
As investigated in Ref.~\cite{Intriligator:1995au}, 
in the SUSY $\text{SU(2)}_H$ gauge theory with three flavours 
(six doublet chiral superfields), the SU(2)$_H$ gauge coupling becomes strong at a confinement scale which 
is denoted by $\Lambda_H$, and below $\Lambda_H$ the low-energy effective theory is described in terms of 
fifteen canonically normalized mesonic composite chiral superfields, 
$H_{ij}\simeq \frac{1}{4\pi \Lambda_H}T_iT_j (i\neq j)$ by using the Naive Dimensional Analysis\cite{NDA}.
The fifteen superfields are summarised in Table \ref{FieldContentHiggs}.
With these mesonic fields, the superpotential in the Higgs sector of the low-energy effective theory is written as 
\begin{align}
	W_{\text{eff}}=&
	{\lambda}
		N\left(H_uH_d+v_0^2\right)
		+
	{\lambda}
		N_{\Phi}\left(\Phi_u\Phi_d+v_{\Phi}^2\right)
		+
	{\lambda}
		N_{\Omega}\left(\Omega_+\Omega_--\zeta\eta+v_{\Omega}^2\right)
		\nonumber\\
		&+
		{\lambda}\left\{
		\zeta H_d\Phi_u
		+\eta H_u\Phi_d
		-\Omega_+H_d\Phi_d
		-\Omega_-H_u\Phi_u
		-NN_{\Phi}N_{\Omega}
\right\}\;,
\end{align}
where the Naive Dimensional Analysis suggests $\lambda\simeq 4\pi$ at the confinement scale $\Lambda_H$.
The relevant soft SUSY breaking terms are given by 
\begin{align}
	\mathcal{L}_H=&
	-m_{H_u}^2H_u^{\dagger}H_u
	-m_{H_d}^2H_d^{\dagger}H_d
	-m_{\Phi_u}^2\Phi_u^{\dagger}\Phi_u
	-m_{\Phi_d}^2\Phi_d^{\dagger}\Phi_d
	\nonumber\\
	&
	-m_N^2 N^*N
	-m_{N_{\Phi}}^2 N_{\Phi}^*N_{\Phi}
	-m_{N_{\Omega}}^2 N_{\Omega}^*N_{\Omega}
	\nonumber\\
	&
	-m_{\Omega_+}^2\Omega_+^*\Omega_+
	-m_{\Omega_-}^2\Omega_-^*\Omega_-
	-m_{\zeta}^2\zeta^*\zeta
	-m_{\eta}^2\eta^*\eta
	\nonumber\\
	&
	-\left\{C\lambda v_0^2 N+C_{\Phi}\lambda v_{\Phi}^2 N_{\Phi} + C_{\Omega}\lambda v_{\Omega}^2 N_{\Omega}+\text{h.c.}\right\}
	\nonumber\\
	&
	-\left\{ B\mu H_uH_d + B_{\Phi}\mu_{\Phi}\Phi_u\Phi_d +B_{\Omega}\mu_{\Omega}(\Omega_+\Omega_-+\zeta\eta)+\text{h.c.}\right\}
	\nonumber\\
	&
	-\lambda\bigl\{ 
		A_NH_uH_dN
		+A_{N_{\Phi}}\Phi_u\Phi_dN_{\Phi}
		+A_{N_{\Omega}}(\Omega_+\Omega_--\eta\zeta)N_{\Omega}
		+A_{\zeta}H_d\Phi_u\zeta 
		\nonumber\\
		&\phantom{-\lambda\bigl\{}+A_{\eta}H_u\Phi_d\eta
		+A_{\Omega_-}H_u\Phi_u\Omega_-
		+A_{\Omega_+}H_d\Phi_d\Omega_+
	+\text{h.c.}\bigr\}
	\nonumber\\
	&-\left\{
	m_{\zeta\eta}^2\eta^*\zeta +\frac{B_{\zeta}^2}{2}\zeta^2
	+\frac{B_{\eta}^2}{2}\eta^2 +\text{h.c.}
	\right\}\;,
\end{align}
where the mass parameters $\mu=\lambda \langle N\rangle$, $\mu_{\Phi}=\lambda \langle N_{\Phi}\rangle$ and $\mu_{\Omega}=\lambda\langle N_{\Omega}\rangle$
are induced after the $Z_2$-even neutral fields $N$, $N_{\Phi}$ and $N_{\Omega}$ get vacuum expectation values (vev's).
As for the field degrees of freedom of the superfields $N_{\Phi}$ and $N_{\Omega}$, 
they are not relevant to the phenomena discussed in this Letter,
therefore we ignore them in the following discussion.
The tree-level Lagrangian for the $Z_2$-even Higgs sector is identical to the one in the nearly-minimal SUSY SM (nMSSM)\cite{nMSSM}.

The matter sector except for the terms relevant to the RH neutrino is almost the same as the one in the minimal SUSY 
SM (MSSM) or the nMSSM, where we assume that the $R$-parity is not broken.
On the other hand, the relevant superpotential terms to the RH neutrino below the confinement scale $\Lambda_H$ are
given by 
\begin{align}
	W_N=&y_N^i N_R^c L_i\Phi_u + h_N^i N_R^c E_i^c\Omega_- +\frac{M_R}{2}N_R^cN_R^c 
	+\frac{\kappa}{2}NN_R^cN_R^c\;,
	\label{eq:Wnu}
\end{align}
where $L_i$ and $E_i^c$ denote the lepton doublets and the charged lepton singlets, respectively.

\section{Mechanisms for baryogenesis (1stOPT), the neutrino masses and the DM}
In the following, we give a  brief review on the mechanisms which are adopted in this model 
for the strong 1stOPT, the neutrino mass generation and the DM.

The condition of the strong 1stOPT, $\varphi_c/T_c\gtrsim 1$, is required for a
successful electroweak baryogenesis scenario.
As discussed in Refs.~\cite{AKS,AKS2,KanemuraSenahaShindouYamada,KanemuraSenahaShindou},
non-decoupling effects of the $Z_2$-odd scalar boson loop can enhance the value 
of $\varphi_c/T_c$.
In our model, this mechanism is adopted.
In order to realize $\varphi_c/T_c>1$, the coupling constant between the SM-like
Higgs boson $h$ and the $Z_2$-odd scalars should be large as $\lambda \simeq 1.8$\cite{KanemuraSenahaShindouYamada},
and masses of the relevant $Z_2$-odd scalars are mainly determined by the contribution from 
the Higgs vev.
For such a large coupling constant as $\lambda \simeq 1.8$, the Landau pole appears at around 
5 TeV which will be identified to the confinement scale $\Lambda_H$.
Above this scale, the theory becomes the SUSY $\text{SU}(2)_H\times Z_2$ gauge theory.

It is known that the same non-decoupling scalar loop effect can also give a significant contribution 
to the triple Higgs boson coupling $\lambda_{hhh}$\cite{KanemuraOkadaSenaha}.
If a charged $Z_2$-odd boson loop gives a significant contribution to $\varphi_c/T_c$, 
it also affects the process of $h\to \gamma\gamma$. 
The HL-LHC with the luminosity of 3000$\text{fb}^{-1}$ is expected to measure the deviation of
$\text{B}(h\to\gamma\gamma)$ from the SM prediction, if it is larger than 10\%\cite{HL-LHC}.
The ILC with $\sqrt{s}=1\text{ TeV}$ with $2.5\text{ ab}^{-1}$ can test the scenario by measuring
the Higgs triple coupling if it deviates in the positive direction 
from the SM prediction as large as $13$\%\cite{ILC}.

\begin{figure}
	\begin{center}
		\begin{tabular}{c}
			\includegraphics[scale=1.0]{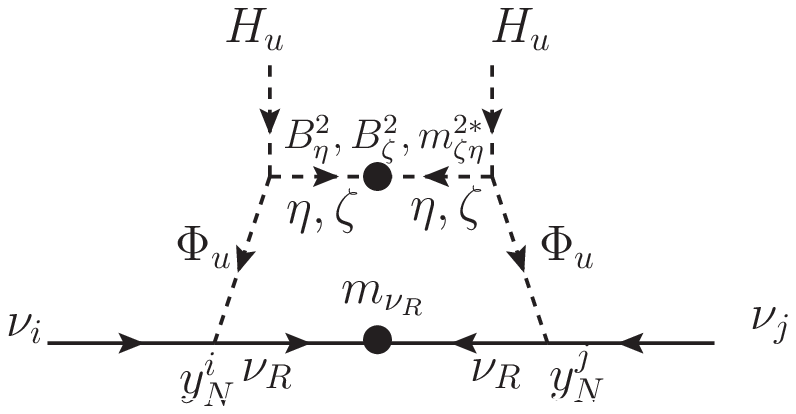}\\
			(I)\\
			\begin{tabular}{cc}
				\includegraphics[scale=0.9]{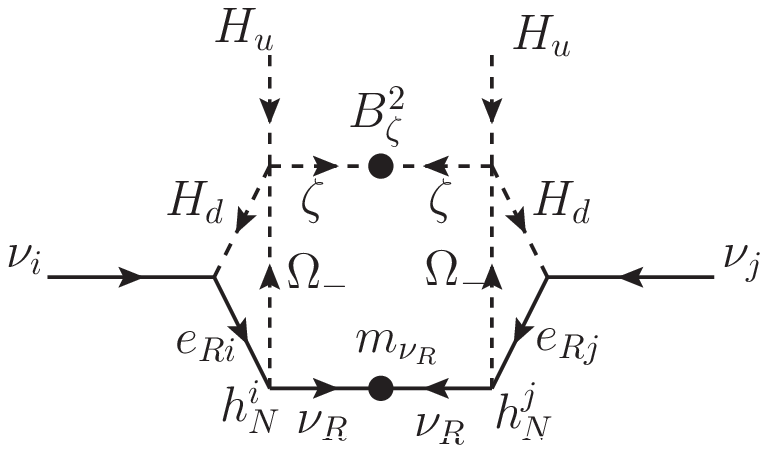}&
				\includegraphics[scale=0.9]{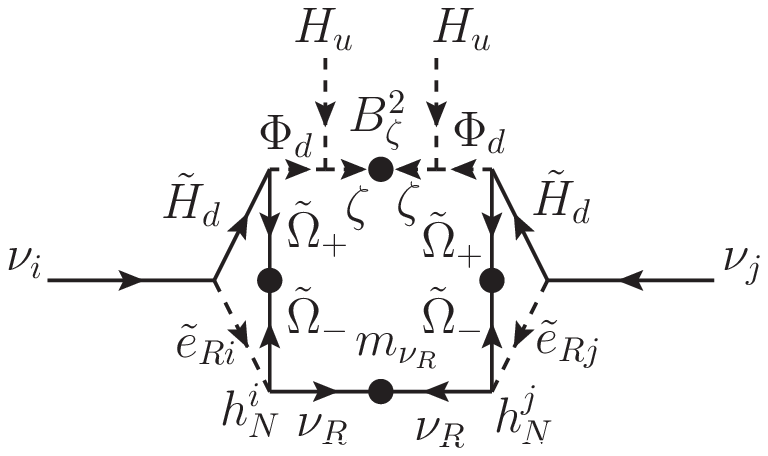}
			\end{tabular}
			\\
			(II)
		\end{tabular}
	\end{center}
	\caption{(I) A one-loop diagram and (II) three-loop diagrams which contribute to the 
	neutrino mass matrix.} 
	\label{numassdiag}
\end{figure}

The neutrino masses in our model are radiatively induced via the hybrid contribution of the one-loop and the three-loop 
diagrams shown in Fig.~\ref{numassdiag}.
The  one-loop diagram and the three-loop diagrams are driven by the coupling constants $y_N^i$ and 
$h_N^i$, respectively, which are independent with each other.
The three-loop contributions are not necessarily suppressed as compared to the one-loop contributions, and 
both the one-loop and the three-loop diagrams can significantly contribute to generating the neutrino masses.
The mass matrix for the neutrino is evaluated as 
\begin{equation}
	(m_{\nu})_{ij}=m_{ij}^{\text{(I)}}+m_{ij}^{(\text{II})}\;,
\end{equation}
where $m_{ij}^{\text{(I)}}$ and $m_{ij}^{\text{(II)}}$ denote 
the one-loop and the three-loop contributions, respectively.
They can be calculated as 
\begin{align}
	m_{ij}^{(\text{I})}=\frac{y_N^{i}y_N^{j}}{(4\pi)^2}
	\left\{
		(O_0^{})^{1\alpha}(O_0^{})^{1\alpha}m_{\nu_R}^{}
	-(O_0^{})^{5\alpha}(O_0^{})^{5\alpha}m_{\nu_R}^{}\right\}\bar{B}_0(m_{\Phi_{\alpha}}^{2},m_{\nu_R}^{2})\;,
\end{align}
and 
\begin{align}
	m_{ij}^{\text{(II)}}=
	&\frac{\hat{\lambda}^4v_u^2y_E^ih_N^{i*}y_E^jh_N^{j*}m_{\nu_R}}{(16\pi^2)^3}\nonumber\\
	&\times
	\sin^4\beta
	(U_{+}^*)_{4\gamma}(U_{+})_{4\gamma}
	(U_{+}^*)_{4\delta}(U_{+})_{4\delta}
	\left\{(O_0)_{2\rho}(O_0)_{2\rho}-(O_0)_{6\rho}(O_0)_{6\rho}\right\}
	\nonumber\\
	&\times 
	F(m_{\nu_R}^2,m_{\Phi_{\rho}}^2;m_{e_i}^2,m_{H^{\pm}}^2,m_{\Phi^{\pm}_{\gamma}}^2;m_{e_j}^2,m_{H^{\pm}}^2,m_{\Phi^{\pm}_{\delta}}^2)
	\nonumber\\
	&+\frac{2\hat{\lambda}^2y_E^ih_N^{i*}y_E^jh_N^{j*}m_{\nu_R}m_{\tilde{\Phi}^{\pm}_{\gamma}}
m_{\tilde{\Phi}^{\pm}_{\delta}}}{(16\pi^2)^3}\nonumber\\
	&\times
	(V_{L}^*)_{2\alpha}(V_{L})_{2\alpha}
	(V_{L}^*)_{2\beta}(V_{L})_{2\beta}
	({U}_L^*)_{2\gamma}({U}_R)_{2\gamma}
	({U}_L^*)_{2\delta}({U}_R)_{2\delta}
	\nonumber\\
	&\times\left\{(O_0)_{3\rho}(O_0)_{3\rho}-(O_0)_{7\rho}(O_0)_{7\rho}\right\}
	F(m_{\nu_R}^2,m_{\Phi_{\rho}}^2;m_{\tilde{\chi}^{\pm}_{\alpha}}^2,m_{\tilde{e}_{Ri}}^2,m_{\tilde{\Phi}^{\pm}_{\gamma}}^2;
	m_{\tilde{\chi}^{\pm}_{\beta}}^2,m_{\tilde{e}_{Rj}}^2,m_{\tilde{\Phi}^{\pm}_{\delta}}^2)
	\;.
	\label{mnu3loop}
\end{align}
In the above expressions, the mixing matrices $O_0$, $U_+$, $U_L$, $U_R$ and
$V_L$
are defined as 
\begin{align}
	&\begin{pmatrix}
		\Phi_u^{\text{even}}\\
		\zeta^{\text{even}}\\
		\Phi_d^{\text{even}}\\
		\eta^{\text{even}}\\
		\Phi_u^{\text{odd}}\\
		\zeta^{\text{odd}}\\
		\Phi_d^{\text{odd}}\\
		\eta^{\text{odd}}\\
	\end{pmatrix}
	=O_0
	\begin{pmatrix}
		\Phi_1\\[2pt]
		\Phi_2\\[2pt]
		\Phi_3\\[2pt]
		\Phi_4\\[2pt]
		\Phi_5\\[2pt]
		\Phi_6\\[2pt]
		\Phi_7\\[2pt]
		\Phi_8\\
	\end{pmatrix}\;,\quad
	\begin{pmatrix}
		\Phi_u^+\\
		\Omega_+\\
		(\Phi_d^-)^*\\
		(\Omega_-)^*
	\end{pmatrix}
	=U_+
	\begin{pmatrix}
		\Phi_1^+\\
		\Phi_2^+\\
		\Phi_3^+\\
		\Phi_4^+\\
	\end{pmatrix}\;,\quad\nonumber\\
	&
	\begin{pmatrix}
		\tilde{\Phi}_d^-\\
		\tilde{\Omega}_-\\
	\end{pmatrix}
	=U_L
	\begin{pmatrix}
		\tilde{\Phi}_{1L}^{-}\\
		\tilde{\Phi}_{2L}^{-}\\
	\end{pmatrix}\;,\quad
	\begin{pmatrix}
		(\tilde{\Phi}_u^+)^*\\
		(\tilde{\Omega}_+)^*\\
	\end{pmatrix}
	=U_R
	\begin{pmatrix}
		\tilde{\Phi}_{1R}^{-}\\
		\tilde{\Phi}_{2R}^{-}\\
	\end{pmatrix}\;,\quad
	\begin{pmatrix}
		\tilde{W}\\
		\tilde{H}_d^-
	\end{pmatrix}
	=V_L
	\begin{pmatrix}
		\tilde{\chi}_{1L}^-\\
		\tilde{\chi}_{2L}^-
	\end{pmatrix}\;,
\end{align}
where the superscript "even" and "odd" denote the CP-even neutral 
scalar component and CP-odd neutral scalar component,
the scalar fields $\Phi_i$ are the mass eigenstates of $Z_2$-odd neutral scalars,  
the scalar fields $\Phi_i^{\pm}$ are the mass eigenstates of $Z_2$-odd charged scalars,  
the fermionic fields $\tilde{\Phi}_{iL}^{-}$ and $\tilde{\Phi}_{iR}^-$ are the 
left-handed and the right-handed components of the mass eigenstates of the
$Z_2$-odd charged fermions, $\tilde{W}$ denotes the wino in the SUSY SM, and $\tilde{\chi}_{iL}^-$ are 
the left-handed component of the mass eigenstates of the $Z_2$-even charginos.
The loop function $\bar{B}_0$ is given by 
\begin{align}
	\bar{B}_0(m_1^2,m_2^2)=-\frac{m_1^2\ln m_1^2-m_2^2\ln m_2^2}{m_1^2-m_2^2}\;,
\end{align}
and the loop function $F$ is given by\cite{AKS,AKS2}
\begin{align}
	&F(M^2,m_{\Phi}^2;m_{\chi 1}^2;m_{\phi 1}^2,m_{\Omega 1}^2;m_{\chi 2}^2;m_{\phi 2}^2,m_{\Omega 2}^2)\nonumber\\
	\nonumber\\
	&=\frac{(16\pi^2)^3}{i}
	\int \frac{d^Dk}{(2\pi)^D}
	\frac{1}{k^2-M^2}
	\frac{1}{k^2-m_{\Phi}^2}
	\int\frac{d^Dp}{(2\pi)^D}\frac{\Fdagger{p}}{p^2-m_{\chi 1}^2}
	\frac{1}{p^2-m_{\phi 1}^2}\frac{1}{(k+p_1)^2-m_{\Omega 1}^2}
	\nonumber\\
	&\phantom{Space}\times \int\frac{d^Dq}{(2\pi)^D}
	\frac{(-\Fdagger{q})}{(-q)^2-m_{\chi 2}^2}
	\frac{1}{(-q)^2-m_{\phi 2}^2}
	\frac{1}{(k+(-q))^2-m_{\Omega 2}^2}\;.
\end{align}
Due to the difference in the flavour structure between 
$m_{ij}^{(\text{I})}$ and $m_{ij}^{(\text{II})}$, two finite mass eigenvalues of light neutrinos 
are induced, even though only one RH neutrino are introduced.

\begin{figure}
	\begin{center}
		\begin{tabular}{cc}
			\includegraphics[scale=1.0]{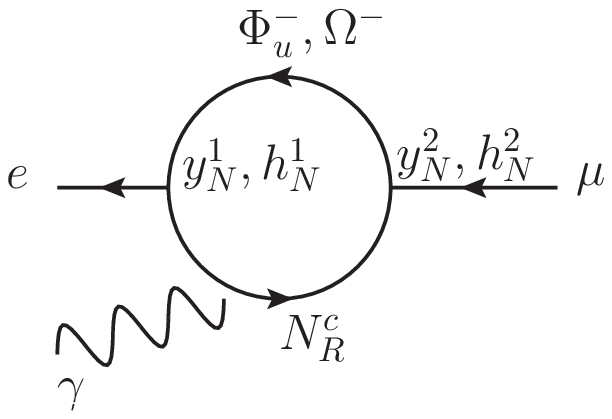}&
			\includegraphics[scale=1.0]{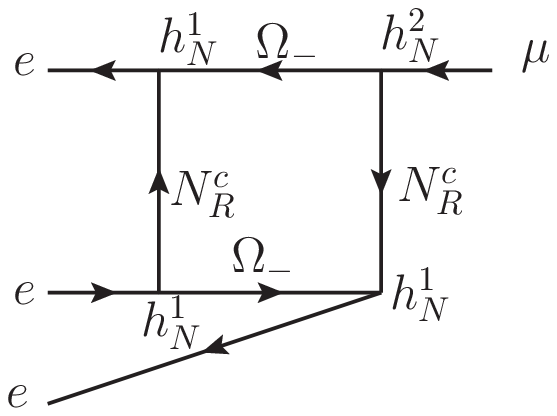}\\
			(a)&(b)
		\end{tabular}
	\end{center}
	\caption{(a) The contribution to the $\mu\to e\gamma$ process 
	and (b) the box contribution to the $\mu\to eee$ process from the $Z_2$-odd particles with the RH neutrino.}
	\label{figLFV}
\end{figure}

The flavour structure in the RH neutrino sector significantly contributes to the 
LFV processes such as $\mu \to e\gamma$ and $\mu\to eee$, which give strong 
constraint on the model parameter space.
The contributions to the $\mu\to e\gamma$ process are from the diagram shown in Fig.~\ref{figLFV}-(a).
The branching ratio $\text{B}(\mu\to eee)$ is suppressed by factor $\frac{\alpha}{4\pi}$ as 
compared to the branching ratio $\text{B}(\mu\to e\gamma)$,
unless the box contribution shown in Fig.~\ref{figLFV}-(b) dominates the branching ratio $\text{B}(\mu\to eee)$.
If the box contribution is significant, the $\mu\to eee$ process gives an independent constraint on the parameter space.
If the MSSM slepton sector has flavour mixing, there will be additional contributions to the LFV.

In the low-energy effective theory of our model, 
two different discrete symmetries: {\it i.e.}, both the $Z_2$-parity and the $R$-parity are 
unbroken. Therefore, there can be three kinds of DM candidates, which are the lightest particles with the 
parity assignments of $(+,-)$, $(-,+)$, and $(-,-)$ 
for $(Z_2\text{-parity},R\text{-parity})$.
The observed value of the thermal relic abundance of DM should be explained by the summation of the relic abundances
of these DM candidates.
If one of the three particle is heavy enough to decay into the other two particles, 
the heaviest one cannot be a DM, and only the other two candidates can compose
the DM relic abundance.
In multi-component DM case\cite{multiDM}, 
not only the pair annihilation processes of each DM candidates but also
the conversion process from one DM particle to the other DM particle can play 
significant role in the evaluation of the relic abundance.

\section{Benchmark scenario}
\begin{figure}[t]
	\begin{center}
		\includegraphics[scale=0.6]{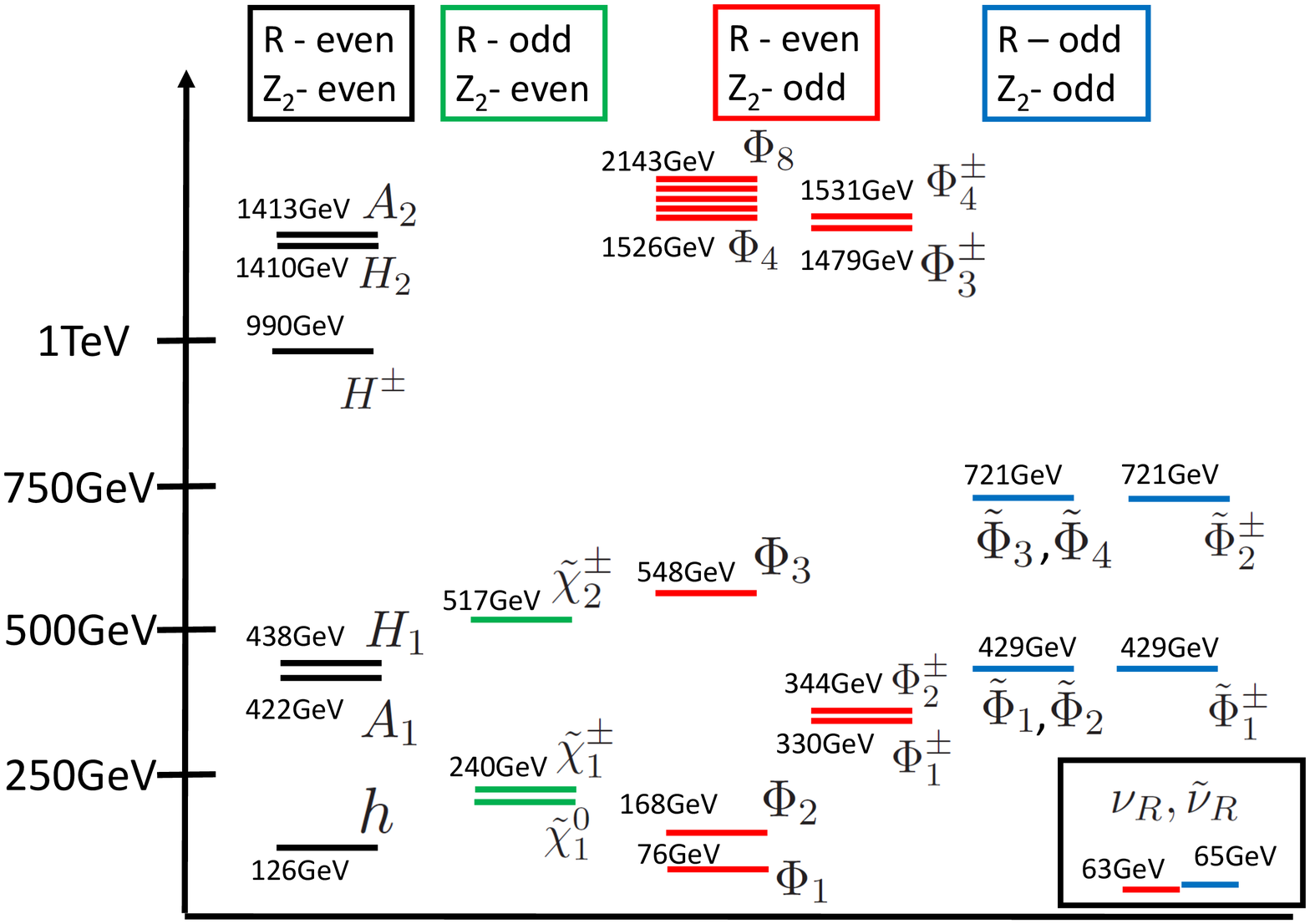}
	\end{center}
	\caption{
		The mass spectrum of the relevant particles in the benchmark scenario 
		given in Table~\ref{Tab:Bench}.
	\label{fig:spectrum}}
\end{figure}

We discuss a benchmark scenario of our model, where
the strong 1stOPT is realized as $\varphi_c/T_c\gtrsim 1$, 
the neutrino oscillation data can be explained with satisfying 
the constraints from the 
LFV processes, and the DM relic abundance can be also reproduced, simultaneously.
The input parameters of the benchmark scenario are listed in Table~\ref{Tab:Bench}.
The predictions in the benchmark scenario are shown in Table~\ref{Tab:Predictions}, and 
the mass spectrum for the relevant particles in the benchmark scenario is shown
in Fig.~\ref{fig:spectrum}.
We here discuss the reason of our choice for the benchmark scenario and its predictions.

\begin{table}[t]
	\caption{The input parameters for the benchmark scenario.
	In the list, $\bar{m}_{\phi_i}^2=m_{\phi_i}^2+|\mu_i|^2$ are taken 
	as input parameters, where $\mu_i=\mu_{\Phi}$ for $\phi_i=\Phi_u,\Phi_d$, and 
	$\mu_i=\mu_{\Omega}$ for $\phi_i=\Omega_{+},\Omega_-,\zeta, \eta$.
	\label{Tab:Bench}}
	\begin{center}
		\renewcommand{\arraystretch}{1.3}
		\begin{tabular}{|c|}\hline
			$\lambda$, $\tan\beta$, and $\mu$-terms \\ \hline
			$\lambda=1.8$ ($\Lambda_H=5$ TeV) \quad $\tan\beta=15$ \quad $\mu=250\;\text{GeV}$
			\quad $\mu_{\Phi}=550\;\text{GeV}$\quad 
			$\mu_{\Omega}=-550\;\text{GeV}$
			\\\hline\hline
			$Z_2$-even Higgs sector\\ \hline
			$m_h=126\;\text{GeV}$\quad $m_{H^{\pm}}=990$\;GeV \quad $m_N^2=(1050\;\text{GeV})^2$ \quad $A_N=2900\;\text{GeV}$
			\\ \hline\hline
			$Z_2$-odd Higgs sector \\ \hline
			$\bar{m}_{\Phi_u}^2=\bar{m}_{\Omega_-}^2=(75\;\text{GeV})^2$\quad 
			$\bar{m}_{\Phi_d}^2=\bar{m}_{\Omega_+}^2=\bar{m}_{\zeta}^2=(1500\;\text{GeV})^2$\quad
			$\bar{m}_{\eta}^2=(2000\;\text{GeV})^2$\\
			$B_{\Phi}=B_{\Omega}=A_{\zeta}=A_{\eta}=A_{\Omega^+}=A_{\Omega^-}=m_{\zeta\eta}^2=0$\quad
			$B_{\zeta}^2=(1400\;\text{GeV})^2$\quad $B_{\eta}^2=(700\;\text{GeV})^2$\\ \hline\hline
			RH neutrino and RH sneutrino sector\\ \hline
			$m_{\nu_R}=63\;\text{GeV}$\quad $m_{\tilde{\nu}_R}=65\;\text{GeV}$\quad $\kappa=0.9$\\
			$y_N=(3.28i, 6.70i, 1.72i)\times 10^{-6}$\quad 
			$h_N=(0,0.227,0.0204)$\\ \hline\hline
			Other SUSY SM parameters\\\hline
			$m_{\tilde{W}}=500\;\text{GeV}$\quad $m_{\tilde{q}}=m_{\tilde{\ell}}=5\;\text{TeV}$\\\hline
		\end{tabular}
	\end{center}
\end{table}

The Lagrangian of the $Z_2$-even Higgs sector is the same as 
the nMSSM.
The $Z_2$-odd sector affects the $Z_2$-even Higgs sector only by the loop effects.
The SM-like Higgs boson mass in our model is estimated as 
\begin{equation}
	m_h^2\simeq m_Z^2\cos^22\beta+\frac{\lambda^2v^2}{2}\sin^22\beta+\delta m_h(\text{loop})\;,
	\label{eq:mh}
\end{equation}
where the $\delta m_h(\text{loop})$ denotes the loop corrections.
If the value of $\tan\beta$ is small, the tree level mass of the SM-like Higgs boson 
becomes too large because of $\lambda\simeq 1.8$, and the measured value $m_h=126$ GeV cannot be reproduced.
Therefore, we take $\tan\beta=15$ in the benchmark scenario.
In this case, the loop correction $\delta m_h(\text{loop})$ plays an important role in 
the determination of the SM-like Higgs boson mass because the second term in Eq.~(\ref{eq:mh}) is 
negligibly small.
The significant loop corrections on the Higgs mass are from the loop contributions of the top and 
stop fields as in the MSSM, as well as from the loop diagrams with $Z_2$-odd fields which has 
large coupling constant with the SM-like Higgs boson.

To realise $\varphi_c/T_c>1$, non-decoupling effect of the $Z_2$-odd particles is 
necessary. The condition $\varphi_c/T_c\gtrsim 1$ requires the large coupling constant 
$\lambda$ as $\lambda \simeq 1.8$ which corresponds to the cut-off scale 
of $\Lambda_H\sim 5$ TeV.
In our model, there are two possible combinations of $Z_2$-odd particles
which give non-decoupling effects on $\varphi_c/T_c$.
One possible way is that $\varphi_c/T_c$ is enhanced by the non-decoupling loop effect
of the scalar component of $\Omega_-$ and the charged scalar component of $\Phi_u$.
This choice is the same as the one discussed in Ref.~\cite{KanemuraSenahaShindouYamada}.
The other is that the enhancement of $\varphi_c/T_c$ is caused by the non-decoupling loop effect 
of the scalar component of $\eta$ and the charged scalar component of $\Phi_u$.
However, for the latter case, the LFV constraint is too severe to avoid the present 
upper bound on $\text{B}(\mu\to e\gamma)$ if only one RH neutrino is introduced. 
In the benchmark scenario, we take the first possibility.
Therefore, the 1stOPT is enhanced as $\varphi_c/T_c=1.3$ by the non-decoupling loop 
contributions of the two charged scalar particles 
$\Phi_1^{-}$ and $\Phi_2^{-}$, whose main components come from the scalar components of 
$(\Phi_u^+)^*$ and $\Omega_-$, respectively.
In this case, the masses of these scalar particles are mainly determined by the vev contributions 
instead of their soft breaking mass parameters.

The non-decoupling effects of the loop contributions by $\Phi_1^{\pm}$ and $\Phi_2^{\pm}$ 
simultaneously affect the predictions on 
both the branching ratio of $h\to \gamma\gamma$ process and
the triple Higgs boson coupling constant $\lambda_{hhh}$\cite{KanemuraOkadaSenaha}.
One can find the minus $20$\% deviation on $\text{B}(h\to \gamma\gamma)$ from the 
SM prediction.
At the present, the LHC data with $\sqrt{s}=7\;\text{TeV}$ and $\sqrt{s}=8\;\text{TeV}$ have determined
the $\text{B}(h\to \gamma\gamma)$ with 50\% accuracy\cite{LHChgamgam},
and the accuracy will be improved to 10\% at the HL-LHC with the 
luminosity of 3000$\text{fb}^{-1}$\cite{HL-LHC}. 
Therefore the model can be tested by measuring the branching ratio of $h\to \gamma\gamma$ at the HL-LHC.
As for the $\lambda_{hhh}$, the plus $20$\% deviation from the SM prediction is predicted, 
and it is testable at the ILC with $\sqrt{s}=1\;\text{TeV}$ with the luminosity 
of $2.5\;\text{ab}^{-1}$ where the $\lambda_{hhh}$ is measured with 13\% 
accuracy\cite{ILC}.

\begin{table}[t]
	\caption{Predictions of the benchmark points given in Table~\ref{Tab:Bench}.
	\label{Tab:Predictions}}
	\begin{center}
		\renewcommand{\arraystretch}{1.3}
		\begin{tabular}{|c|}\hline
			Non-decoupling effects\\\hline
			$\varphi_c/T_c=1.3$\quad $\lambda_{hhh}/\lambda_{hhh}|_{\text{SM}}=1.2$\quad 
			$\text{B}(h\to \gamma\gamma)/\text{B}(h\to \gamma\gamma)|_{\text{SM}}=0.78$\\ \hline\hline
			Neutrino masses and the mixing angles\\ \hline
			$(m_1, m_2, m_3)=(0, 0.0084\;\text{eV}, 0.0050\;\text{eV})$\quad
			$\sin^2\theta_{12}=0.32$\quad 
			$\sin^2\theta_{23}=0.50$\quad 
			$|\sin\theta_{13}|=0.14$\\ \hline\hline
			LFV processes\\ \hline
			$\text{B}(\mu\to e\gamma)=3.6\times 10^{-13}$\quad 
			$\text{B}(\mu\to eee)=5.6\times 10^{-16}$\\ \hline\hline
			Relic abundance of the DM\\ \hline
			$\Omega_{\nu_R}h^2=0.055$\quad 
			$\Omega_{\tilde{\nu}_R}h^2=0.065$\quad 
			$\Omega_{\text{DM}}=\Omega_{\nu_R}h^2+\Omega_{\tilde{\nu}_R}h^2=0.12$\\ \hline
		\end{tabular}
	\end{center}
	\end{table}

Two finite mass eigenvalues can be obtained with only one RH neutrino. 
In the benchmark scenario, the solar neutrino mass difference is mainly induced by the one-loop 
contribution shown in Fig.~\ref{numassdiag}-(I), and the atmospheric neutrino mass difference is dominated by the three-loop 
contributions shown in Fig.~\ref{numassdiag}-(II).
As shown in Table~\ref{Tab:Predictions}, 
the predicted mass eigenvalues and the mixing angles are consistent with their allowed region 
which is obtained from the global fitting analysis of the neutrino oscillation data as\cite{nufit}
\begin{align}
	&2.28\times 10^{-3}\;\text{eV}^2<|m_3^2-m_1^2|<2.70\times 10^{-3}\;\text{eV}^2\;,\quad\nonumber\\
	&7.0\times 10^{-5}\;\text{eV}^2<m_2^2-m_1^2<8.1\times 10^{-5}\text{eV}^2,\nonumber\\
&0.27<\sin^2\theta_{12}<0.34\;,\quad
0.34<\sin^2\theta_{23}<0.67\;,\quad
0.016<\sin^2\theta_{13}<0.030\;.
\end{align}
The light neutrino mass pattern in our benchmark is the normal hierarchy ($m_1<m_2<m_3$). 
It is difficult to reproduce the inverted hierarchical pattern ($m_3<m_1<m_2$) with satisfying the LFV constraint 
when only one RH neutrino is introduced.

The experimental upper bound on the branching ratio $\text{B}(\mu\to e\gamma)$ gives a severe 
constraint on the parameter space. 
In the benchmark scenario, 
though the contribution to the $\mu\to e\gamma$ process is suppressed to some extent by taking $h_N^1=0$, 
the predicted value of the branching ratio of $\mu\to e\gamma$ as $\text{B}(\mu\to e\gamma)=3.6\times 10^{-13}$ 
is just below the present upper limit such as $\text{B}(\mu\to e\gamma)\simeq 5.7\times 10^{-13}$, which is given by the 
MEG experiment\cite{MEG}.
The box diagram contribution to $\mu\to eee$ in the benchmark scenario 
is negligible compared to the penguin and dipole contributions because of $h_N^1=0$.
Therefore, the predicted branching ratio of $\mu\to eee$ easily satisfies the experimental upper limit 
such as $\text{B}(\mu\to eee)\simeq 10^{-12}$\cite{mueee}.

\begin{figure}[t]
	\begin{center}
		\begin{tabular}{cc}
			\includegraphics[scale=1.0]{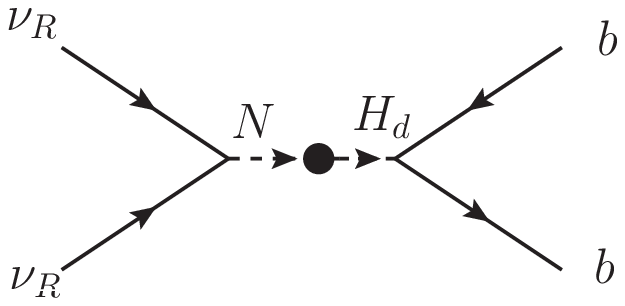}&
			\includegraphics[scale=1.0]{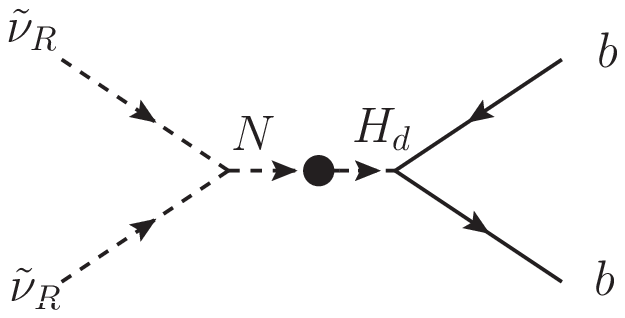}\\
			(a)&(b)\\
			\includegraphics[scale=1.0]{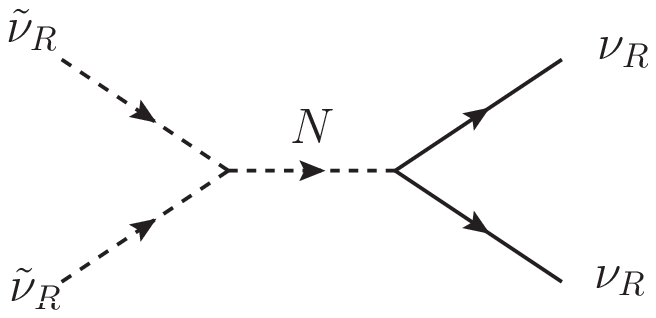}&\\
			(c)&
		\end{tabular}
	\end{center}
	\caption{The main annihilation processes of (a) the RH neutrino and (b) the RH sneutrino and 
	(c) the conversion process from the RH neutrino pair to the RH sneutrino pair.}
	\label{annihDM}
\end{figure}

\begin{figure}[t]
	\begin{center}
		\begin{tabular}{cc}
			\includegraphics[scale=0.30]{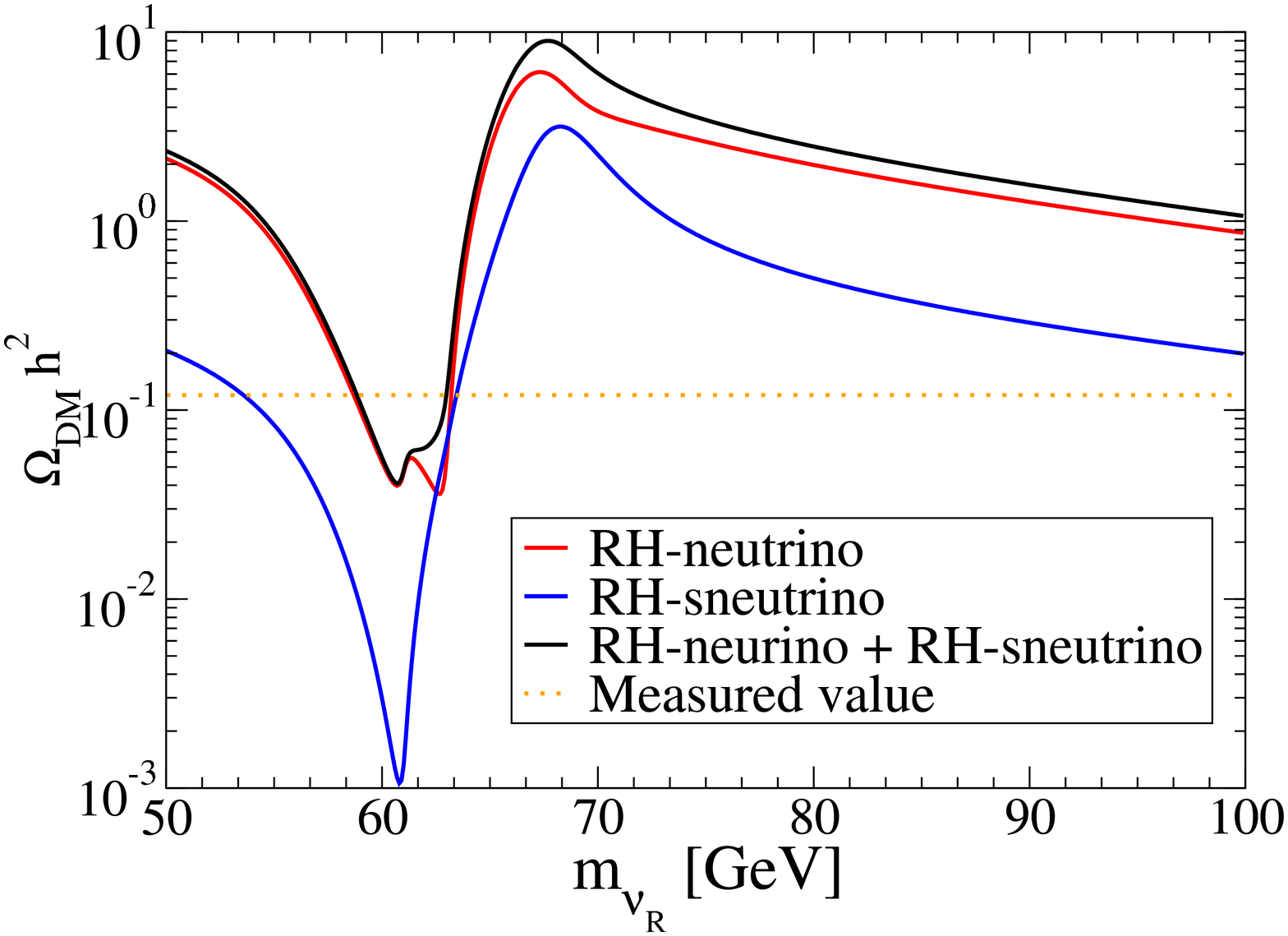}&
			\includegraphics[scale=0.30]{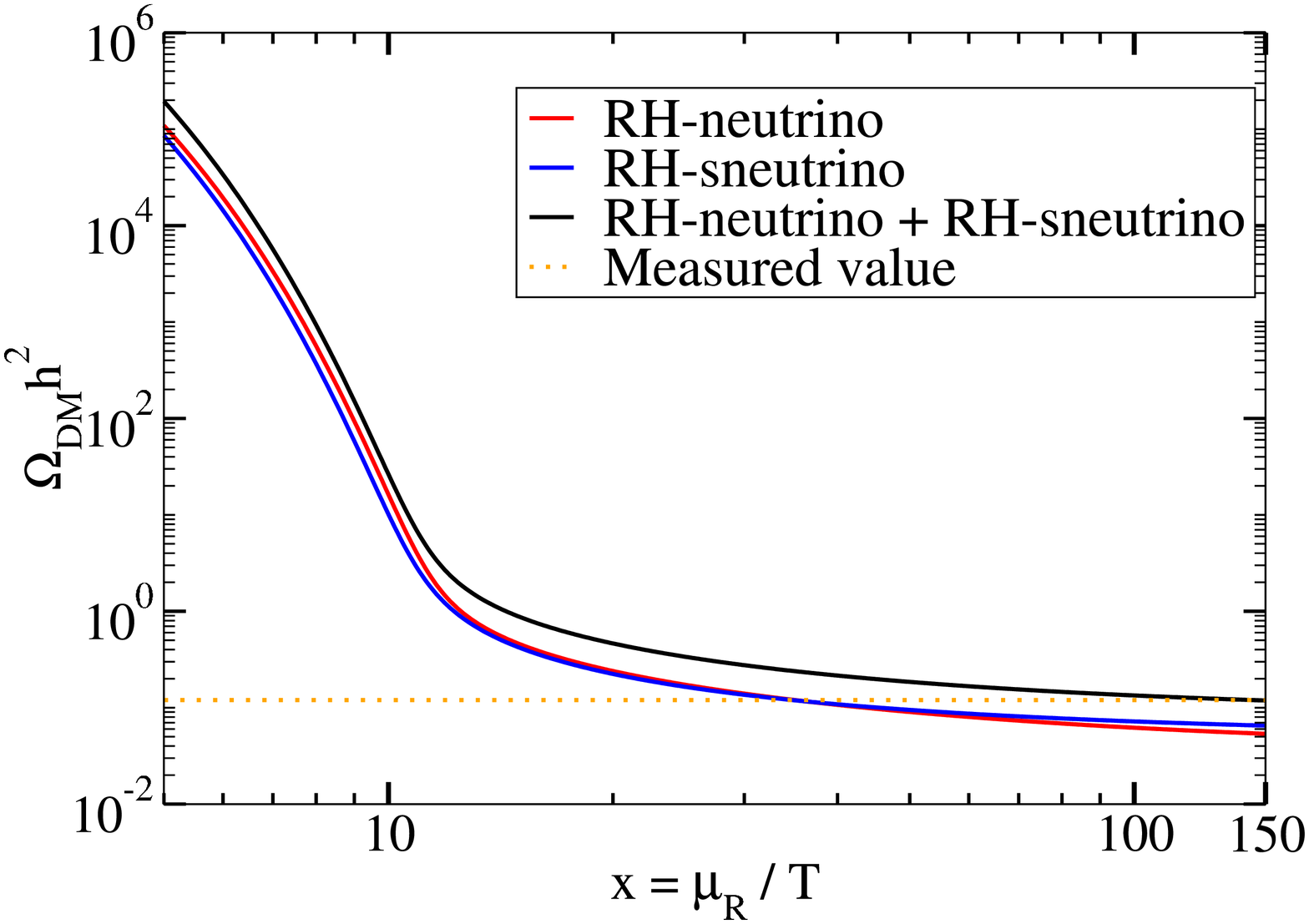}\\
		(a)&(b)
	\end{tabular}
	\end{center}
	\caption{
		(a) The prediction on the thermal relic densities of the RH neutrino 
		and the RH sneutrino as functions of the RH neutrino mass $m_{\nu_R}$.
		The sneutrino mass is taken as $m_{\tilde{\nu}_R}=m_{\nu_R}+2\text{GeV}$.
		The other parameters are the same as ones in Table~\ref{Tab:Bench}.
		(b) The behaviour of the thermal relic abundances in the benchmark scenario:
		{\it i.e.}, the mass of the RH neutrino and the RH sneutrino are fixed as 
		$m_{\nu_R}=63\;\text{GeV}$ and $m_{\tilde{\nu}_R}=65\;\text{GeV}$. 
	\label{fig:DM}}
\end{figure}

There are three DM candidates in our model; {\it i.e.}, the 
lightest particles with the parity assignments of $(+,-)$, $(-,+)$, and $(-,-)$ 
for the $(Z_2\text{-parity},R\text{-parity})$.
In our benchmark scenario, the lightest  $(+,-)$, $(-,+)$, and $(-,-)$ 
particles are identical to the lightest $Z_2$-even neutralino, the RH neutrino
and the RH sneutrino, respectively.
One may consider another possibility for the lightest $(-,+)$ and $(+,-)$ particles 
such as $\Phi_1$ and $\tilde{\Phi}_1$. However, different from the RH neutrino and 
RH sneutrino, the other $Z_2$-odd particle have gauge interactions in addition to the large
coupling constant with the SM-like Higgs boson. Therefore, the scattering cross section with the proton 
is too large to avoid the constraint from the direct detection experiments such as 
the XENON100 experiment\cite{XENON100} and the LUX experiment\cite{LUX}. 
In the benchmark scenario, 
the lightest $Z_2$-even neutralino $\tilde{\chi}_1^0$ is heavy enough to 
decay into the RH neutrino and the RH sneutrino, and it cannot be a DM candidate.
Consequently, there are only two DM candidate; {\it i.e.}, the RH neutrino and 
the RH sneutrino.
In Section~\ref{Sec:DManalysis}, we show the brief discussion of the numerical analysis of the relic abundance 
in this two component DM system.
The annihilation and the conversion processes of the RH neutrino 
and the RH sneutrino are dominated by 
the exchange of the $Z_2$-even singlet scalar $N$ which mixes to 
the SM-like Higgs boson.
The diagrams of the annihilation processes are shown in Fig.~\ref{annihDM}.
As shown in Fig.~\ref{fig:DM}, in order to reproduce the observed DM relic abundance 
$\Omega_{\text{DM}}h^2=\Omega_{\nu_R}h^2+\Omega_{\tilde{\nu}_R}h^2
\simeq 0.12$ \cite{WMAP}, 
the masses of the RH neutrino and the RH sneutrino should be about one half of the SM-like 
Higgs boson mass, $m_{\nu_R}\simeq m_{\tilde{\nu}_R}\simeq m_h/2$.
In this case, the effect of the s-channel resonance 
can enhance the annihilation processes enough to reproduce 
the observed relic abundance of DM.
The coupling between the DM particles and the SM-like Higgs boson is determined
by the combination of the coupling constant $\kappa$ and the mixing angles
in the $Z_2$-even and CP-even neutral scalar sector.
In order to enhance the annihilation process in this way, 
the mixing among the scalar components of $N$ and 
$H_d$ has to be large, 
and the SM-like Higgs boson should contain the non-negligible 
component from the scalar component of the singlet $N$.

Since both the RH neutrino and the RH sneutrino are gauge singlet fields, 
they scatter off the proton only through the 
Higgs exchange diagram, and the scattering cross section with the proton is suppressed by 
the Yukawa coupling constant of light quarks such as $u$, $d$, and $s$.
Then the cross section is far below the current limits by direct detection experiments\cite{XENON100,LUX}.

If the $Z_2$-even neutralino is too light to decay into the RH neutrino and the RH sneutrino, 
the neutralino $\tilde{\chi}_1^0$ can also be a DM candidate in addition to the RH neutrino and the RH sneutrino.
In this case, some additional mechanism to accelerate the annihilation of $\tilde{\chi}_1^0$ is necessary 
to reproduce the observed relic abundance of DM; {\it e.g.}, co-annihilation with stau and so on.

Our simple benchmark scenario given in Table~\ref{Tab:Bench} can explain 
the DM relic abundance, 
the neutrino oscillation data with satisfying the experimental bound from the LFV
process and with retaining the strong 1stOPT for 
successful electroweak baryogenesis by introducing only one RH neutrino superfield.

\section{Analysis of the DM relic abundance}
\label{Sec:DManalysis}
We here briefly show how the relic abundance is numerically evaluated 
with the two DM particles; {\it i.e.}, the RH neutrino and the RH sneutrino.
The relic abundance of DM in this scenario is the summation of the 
relic abundances of the RH neutrino and the RH sneutrino.
These relic abundances are evaluated by using the coupled Boltzmann equations as
\begin{align}
\frac{d Y}{d x}
= &
0.264 g_*^{1/2}\left(\frac{\mu_R M_{P}}{x^2} \right)
\nonumber\\
&
\times
\left\{
	- \langle\sigma_{\nu} v\rangle 
	\left( Y^2 - Y_{\text{eq}}^2 \right)
- \langle \sigma_{\nu\tilde{\nu}} v\rangle 
\left( Y^2 - \tilde{Y}^2 \frac{Y_{\text{eq}}^2}{\tilde{Y}_{\text{eq}}^2} \right)
	+ \langle \sigma_{\tilde{\nu}\nu}v\rangle 
	\left( \tilde{Y}^2 - Y^2 \frac{\tilde{Y}_{\text{eq}}^2}{Y_{\text{eq}}^2} \right)
\right\}\;,\nonumber \\
\frac{d \tilde{Y}}{d x}
= &
0.264 g_*^{1/2}\left(\frac{\mu_R M_{P}}{x^2} \right)
\nonumber\\
&
\times
\left\{
	- \langle\sigma_{\tilde{\nu}} v\rangle 
	\left( \tilde{Y}^2 - \tilde{Y}_{\text{eq}}^2 \right)
	- \langle \sigma_{\tilde{\nu}\nu} v\rangle 
	\left( \tilde{Y}^2 - Y^2 \frac{\tilde{Y}_{\text{eq}}^2}{{Y}_{\text{eq}}^2} \right)
	+ \langle \sigma_{{\nu}\tilde{\nu}}v\rangle 
	\left( \tilde{Y}^2 - Y^2 \frac{\tilde{Y}_{\text{eq}}^2}{Y_{\text{eq}}^2} \right)
\right\}\;.
\end{align}
In the above expressions, 
$Y$ and $\tilde{Y}$ denote the ratio of the particle number density to the entropy density for 
the RH neutrino and the RH sneutrino, respectively.
$Y_{\text{eq}}$ and $\tilde{Y}_{\text{eq}}$ are the equilibrium numbers for the $Y$ and $\tilde{Y}$, 
$x$ is the dimensionless inverse temperature $x=\frac{\mu_R}{T}$ with $\mu_R$ being 
the reduced mass of the two component system as $\mu_R^{-1}=m_{\nu_R}^{-1}+m_{\tilde{\nu}_R}^{-1}$, 
$g_*^{1/2}$ is a parameter for the effective degrees of freedom in the thermal equilibrium, and 
$M_P$ is the Planck mass.
In the thermal averaged cross sections $\langle \sigma v\rangle$, 
the cross sections $\sigma_{\nu}$, $\sigma_{\tilde{\nu}}$, 
$\sigma_{\nu\tilde{\nu}}$, and $\sigma_{\tilde{\nu}\nu}$ are 
relevant to the processes such as $\nu_R\nu_R\to XX$ ($X$ denotes a generic SM fermion particles.), 
$\tilde{\nu}_R\tilde{\nu}_R\to XX$, 
$\nu_R\nu_R\to \tilde{\nu}_R\tilde{\nu}_R$ and 
$\tilde{\nu}_R\tilde{\nu}_R\to \nu_R\nu_R$, respectively.
In this benchmark scenario, $\sigma_{\nu\tilde{\nu}}$ is kinematically suppressed.
The relic densities of the RH neutrino and the RH sneutrino are evaluated from the frozen out values of $Y$ and $\tilde{Y}$ as 
\begin{equation}
	\Omega_{\nu_R}h^2=2.74\times 10^{8}\left(\frac{m_{\nu_R}}{1 \text{GeV}}\right) Y\;,\quad 
	\Omega_{\tilde{\nu}_R}h^2=2.74\times 10^{8}\left(\frac{m_{\tilde{\nu}_R}}{1 \text{GeV}}\right) \tilde{Y}\;.
\end{equation}
The numerical behaviour of the thermal relic abundance of the RH neutrino and the RH sneutrino 
in the benchmark scenario is shown in Fig.~\ref{fig:DM}-(b).

\section{Discussion}
\begin{table}
	\caption{The deviations in the coupling constants 
	from the SM predictions in the benchmark scenario.\label{Tab:HiggsFingerPrint}}
	\begin{center}
		\renewcommand{\arraystretch}{1.3}
	\begin{tabular}{|c|c|c|c|c|c|c|} \hline
		$\kappa_W$&$\kappa_Z$&$\kappa_u$&$\kappa_d$&$\kappa_\ell$&
		$\kappa_\gamma$&$\lambda_{hhh}/\lambda_{hhh}^{\text{SM}}$\\ \hline
		$0.990$&$0.990$&$0.990$&$0.978$&$0.978$&$0.88$&$1.2$\\ \hline
		\end{tabular}
\end{center}
\end{table}

For electroweak baryogenesis, we have focused on the strong 1stOPT which is one of the necessary conditions 
for successful baryogenesis. 
Towards a complete analysis of generation of the BAU, the CP violating phases should also be
taken into account.
Since it is known that the CP violation in the SM is not enough for the successful 
baryogenesis\cite{BGinSM},
new CP violating source is required to be introduced.
In the SUSY model, several new CP violating phases can be introduced, some of which 
can contribute to the baryogenesis\cite{CPinSUSY}.
With such CP phases, the BAU in the electroweak baryogenesis scenario is numerically 
evaluated in the MSSM\cite{CPinMSSM}.
In our model, by introducing CP phase to the model in the similar way to the case of the MSSM, 
we expect to reproduce the measured amount of the BAU, if the 1stOPT is strong enough.
However, it should be carefully checked if introducing such a CP phase does not 
conflict with the experimental constraints as the bounds on the neutron 
electric dipole moment and so on\cite{CPinMSSM}. 
The complete analysis for getting the BAU in our model will be performed elsewhere.

Let us discuss the testability of our model.
In the benchmark scenario, $Z_2$-odd scalars $H_1$ and $A_1$ are rather light as 
$m_{H_1}=438\;\text{GeV}$ and $m_{A_1}=422\;\text{GeV}$. Such masses for $\tan\beta=15$ can be easily 
searched at the LHC with $\sqrt{s}=14\;\text{TeV}$\cite{2HDMLHC}.
When they are discovered, 
they may look like the heavy Higgs and the CP-odd Higgs in 
the MSSM or the two Higgs doublet model.
On the other hand, the $Z_2$-even charged Higgs is not degenerate to the $H_1$ and $A_1$ in the 
benchmark scenario as $m_{H^{\pm}}=990\;\text{GeV}$. This mass spectrum is quite different 
from the MSSM in which it is known a mass relation is satisfied $m_{H^{\pm}}^2=m_A^2+m_W^2$
for the charged Higgs mass $m_{H^{\pm}}$ and the CP-odd Higgs mass $m_A$. Therefore 
our model can be distinguished from the MSSM.
In addition, their property will be precisely measured at the ILC with $\sqrt{s}=1\;\text{TeV}$.
Both $H_1$ and $A_1$ in the benchmark scenario are mixture 
of the doublet and the singlet. The precision 
measurements of these heavy state; {\it e.g.}, coupling measurement with bottom quarks and 
tau leptons, also provide enough information to distinguish our model from the MSSM,
the two Higgs doublet model and so on. 

Since such a mass spectrum and properties from the mixture with the singlet state are 
found in the nMSSM too, it is hard to distinguish our model from the nMSSM by these 
measurements only.
However, in our model, the $Z_2$-odd sector affects the $Z_2$-even Higgs sector 
through the non-decoupling loop effect, which will be explored by precision measurements
of the SM-like Higgs boson at future collider experiments.
Table~\ref{Tab:HiggsFingerPrint} shows the deviations from the SM prediction in the 
coupling constants of the SM-like Higgs boson. The deviations are parametrised by 
the scale factors
$\kappa_{\phi}\equiv g_{h\phi\phi}/g_{h\phi\phi}^{\text{SM}}$ for $h\phi^{(*)}\phi$
couplings ($\phi=Z,W,u,d,\ell, \gamma$).
The deviations in $\kappa_{W}$, $\kappa_{Z}$, $\kappa_{u}$, $\kappa_{d}$ and
$\kappa_{\ell}$ mainly originate from the mixture between the SM-like Higgs boson and the singlet 
scalar component of $N$, while the deviations in $\kappa_{\gamma}$ and 
the triple Higgs boson coupling $\lambda_{hhh}/\lambda_{hhh}^{\text{SM}}$ are 
caused by the non-decoupling effect of $Z_2$-odd particles.
Therefore, the deviations in $\kappa_{\gamma}$ and $\lambda_{hhh}/\lambda_{hhh}^{\text{SM}}$ can distinguish our model 
from the nMSSM. It is expected that 
the deviation in $\kappa_{\gamma}$ can be tested with a few percent accuracy at the 
HL-LHC with the luminosity of 3000$\text{fb}^{-1}$\cite{HL-LHC}. 
For $\lambda_{hhh}$, the ILC with $\sqrt{s}=1\;\text{TeV}$ 
with the luminosity of $2.5\text{ ab}^{-1}$ can measure the positive deviation 
at most the $13\%$ accuracy\cite{ILC}.
Therefore, our model can be tested by measuring the self coupling constant of the SM-like Higgs boson.

Even if $H_1$ and $A_1$ are heavier so that they are not discovered at the 
LHC with $\sqrt{s}=14\;\text{TeV}$, the precision measurements of the SM-like Higgs boson 
are very powerful tool to explore the framework of our model.
We can consider a benchmark case with much heavier $H_1$ and $A_1$, where
LFV constraint becomes more severe, but it is avoidable by introducing the second RH neutrino superfield.
In such a benchmark with heavier $H_1$ and $A_1$, a few percent of deviations can appear in
$\kappa_{W}$, $\kappa_{Z}$, $\kappa_{u}$, 
$\kappa_{d}$ and $\kappa_{\ell}$ 
caused by the mixture of the SM-like Higgs boson and the singlet scalar $N$.
Precision measurements of these scale factors give us a strong hint to distinguish our model from the MSSM.

The existence of light $Z_2$-odd particles characterize our benchmark scenario so that the signals in the 
direct search of the $Z_2$-odd particles are very important.
In the literature\cite{collider_inert}, collider phenomenology of $Z_2$-odd doublet scalars have been 
discussed.
In a specific case, the $Z_2$-odd scalars might be discovered at the LHC by using the cascade decays of 
heavier particles.
However, 
in general, it is not easy to discover them at the LHC because 
these $Z_2$-odd particles are colour singlet particles.
On the other hand, the ILC is a strong tool for not only discovering them but also for determining 
their masses and quantum numbers.
As discussed in Ref.~\cite{AokiKanemuraYokoya}, the mass of a neutral $Z_2$-odd doublet-like scalar 
can be determined in more than 2 GeV accuracy, and a $Z_2$-odd charged scalar mass can be measured 
in a few GeV accuracy at the ILC with $\sqrt{s}=250$ GeV.

In our model, significant size of the LFV is unavoidable, because the origin of the 
neutrino mass in our model is at the TeV scale.
Actually, in the benchmark scenario, the prediction on the branching ratio of 
$\mu\to e\gamma$ is just below the present upper limit.
Therefore, 
a signal of $\mu\to e\gamma$ is strongly expected to be found
in a future experiment such as an upgrade version of the MEG experiment 
\cite{MEGupgrade}, 
whose sensitivity on the $\mu\to e\gamma$  will reach $\text{B}(\mu\to e\gamma)<10^{-14}$.

\section{Summary}
We propose a simple model to explain the problems which cannot be 
explained in the SM; {\it i.e.},
tiny neutrino mass, DM and
baryogenesis.
The model is based on the idea that the extended Higgs sector appears 
as a low-energy effective theory of a SUSY gauge theory 
with confinement.
We have considered the $\text{SU(2)}_H$ gauge symmetry with three flavours
of fundamental representations and a new discrete $Z_2$ symmetry.
A $Z_2$-odd RH neutrino superfield is also introduced.
In the low-energy effective theory, SUSY extended Higgs sector appears, 
where there are several $Z_2$-odd composite superfields.
When the confinement scale is of the order of ten TeV,
electroweak phase transition can be sufficiently of first order for 
successful electroweak baryogenesis by the non-decoupling effect of the $Z_2$-odd 
particles by the non-decoupling effect of the $Z_2$-odd 
particles.
In addition to the lightest $R$-parity odd DM candidate, 
the lightest $Z_2$-odd particle can be a new candidate for DM.
Neutrino masses and mixings can be explained by the quantum effects of 
$Z_2$-odd fields via the one-loop and three-loop diagrams.
We have found a simple benchmark scenario of the model, where all the constraints from 
neutrino, DM, LFV and LHC data are satisfied.
We have also discussed its testability at future collider experiments and LFV experiments.

\section*{Acknowledgement}
We would like to thank Toshifumi Yamada for useful discussions.
This work was supported in part by Grant-in-Aid for Scientific Research, Japan Society for the 
Promotion of Science (JSPS) and Ministry of Education, Culture, Sports, 
Science and Technology,
Nos. 22244031 (S.K.), 23104006 (S.K.), 23104011 (T.S.) and 24340046 (S.K. and T.S.). 
The work of N.M. was supported in part by the Sasakawa Scientific Research Grant from the Japan 
Science Society.

\end{document}